\typein[\sorb]{ big or little (b/l)?}
\newlength{\absize}
\input prepictex
\input pictex
\input postpictex
\def\square{\beginpicture\setcoordinatesystem units <\unitlength,\unitlength>
\putrule from -5 -5 to -5 5
\putrule from 5 -5 to 5 5
\putrule from -5 -5 to 5 -5
\putrule from -5 5 to 5 5
\endpicture}
\if l\sorb \documentstyle{article}
\renewcommand{\baselinestretch}{1.5}
\renewcommand{\arraystretch}{1.5}
\begin{document}
\date{}
\pagestyle{empty}
\thispagestyle{empty}
\renewcommand{\thefootnote}{\fnsymbol{footnote}}
\newcommand{\starttext}{\newpage\normalsize
\pagestyle{plain}
\setlength{\baselineskip}{4ex}\par
\twocolumn\setcounter{footnote}{0}
\renewcommand{\thefootnote}{\arabic{footnote}}
}
\else
\documentstyle[12pt]{article}
\setlength{\absize}{6in}
\setlength{\topmargin}{-.5in}
\setlength{\oddsidemargin}{-.3in}
\setlength{\evensidemargin}{-.3in}
\setlength{\textheight}{9in}
\setlength{\textwidth}{7in}
\renewcommand{\baselinestretch}{1.5}
\renewcommand{\arraystretch}{1.5}
\setlength{\footnotesep}{\baselinestretch\baselineskip}
\begin{document}
\thispagestyle{empty}
\pagestyle{empty}
\renewcommand{\thefootnote}{\fnsymbol{footnote}}
\newcommand{\starttext}{\newpage\normalsize
\pagestyle{plain}
\setlength{\baselineskip}{4ex}\par
\setcounter{footnote}{0}
\renewcommand{\thefootnote}{\arabic{footnote}}
}
\fi

\setcounter{bottomnumber}{2}
\setcounter{topnumber}{3}
\setcounter{totalnumber}{4}
\newcommand{\figsize}{\small}
\renewcommand{\bottomfraction}{1}
\renewcommand{\topfraction}{1}
\renewcommand{\textfraction}{0}
\newdimen\tdim
\tdim=1.3\unitlength
\newcommand{\preprint}[1]{\begin{flushright}
\setlength{\baselineskip}{3ex}#1\end{flushright}}
\renewcommand{\title}[1]{\begin{center}\LARGE
#1\end{center}\par}
\renewcommand{\author}[1]{\vspace{2ex}{\Large\begin{center}
\setlength{\baselineskip}{3ex}#1\par\end{center}}}
\renewcommand{\thanks}[1]{\footnote{#1}}
\renewcommand{\abstract}[1]{\vspace{2ex}\normalsize\begin{center}
\centerline{\bf Abstract}\par\vspace{2ex}\parbox{\absize}{#1
\setlength{\baselineskip}{2.5ex}\par}
\end{center}}
\newcommand{\lnc}{large $N_c$}
\newcommand{\n}[1]{N_c=#1}
\preprint{\#HUTP-93/A032\\ 10/93}
\title{On Spin Independence in Large $N_c$ Baryons\thanks{Research
supported in part by the National Science Foundation under Grant
\#PHY-9218167.}\thanks{Research supported in part by the Texas National
Research Laboratory Commission, under Grant \#RGFY9206.}}
\author{Chris Carone, Howard Georgi and Sam Osofsky\\
Lyman Laboratory of Physics \\
Harvard University \\
Cambridge, MA 02138 \\
\vspace{1ex}}
\date{}
\abstract{We argue directly from Witten's analysis of large $N_c$ baryons that
the structure of the $s$-wave low-spin baryon states in QCD becomes
spin-independent as $N_c\rightarrow\infty$. This property leads to
$SU(6)$-like behavior of static matrix elements, such as the axial-vector
current matrix elements recently studied by Dashen, Manohar and Jenkins. Our
analysis suggests a method for applying large $N_c$ results for $N_c=3$, even
though the baryon states for large $N_c$ are very different.}
\starttext

\subsection*{\label{intro}Introduction}

The classic paper by Witten on baryons in the \lnc\ approximation shows that
low-lying \lnc\ baryons can be described by a Hartree wave-function with all
(or almost all, for low-lying excited states) quarks in the same ground-state
wave function, bound in a potential produced by all the other
quarks.~\cite{witten} One sentence in \cite{witten} hints that the spin
structure of \lnc\ baryonic bound states of light quarks may be an interesting
thing to study. Witten notes that while spin-orbit coupling will seriously
deform the high-spin baryons away from an $s$-wave ground state, the low-spin,
ground state baryons may not be deformed. In this note, we attempt to make
this notion precise. We will argue that spin-independence of \lnc\ baryons
emerges as an approximate symmetry of a rather unusual type. Spin dependent
terms vanish as $N_c\rightarrow\infty$ when the baryon spin is held fixed.
Thus the states of low spin in the baryon multiplet are spin-independent,
while the states with spin of order $N_c/2$ are seriously modified by
spin-spin and spin-orbit interactions. We will then analyze the matrix
elements of operators in these baryon states and show that for the ground
states, the matrix elements have the structure suggested by spin-flavor
symmetry arguments. Applied to the axial-vector current, this yields the
results that Dashen and Manohar~\cite{dandm} and Jenkins~\cite{jenkins}
obtained by studying pion-baryon scattering. However, we see that the result
is far more general, depending only \lnc, not on chiral symmetry.

One difficulty that arises in applying \lnc\ ideas to baryons in our world
is that baryon states for \lnc\ look nothing like the baryons for $N_c=3$. It
is purely coincidental that for only two flavors of quarks, the low lying
states for any odd $N_c>3$ have the same quantum numbers as the states for
$\n3$. For more than two flavors, the quantum numbers of the low-lying states
look entirely different. It is thus important to extract results from a \lnc\
analysis in a form that can be unambiguously applied to $\n3$. It seems clear
that you should not directly compute properties of the \lnc\ baryon states and
then just take over the results to $\n3$. What we will suggest is to formulate
the result as a sum over quark states, without regard to the value of $N_c$.
This approach fits in nicely with the rest of our analysis.

We will work entirely within the Hartree picture of \lnc\ baryons. However, to
get started, we must know what spin states to consider. We begin by
considering heavy quarks, as discussed explicitly by Witten. In this case, we
know what the Hartree Hamiltonian looks like. The interactions are
approximately spin independent simply because the quark masses are large and
the baryons are nonrelativistic bound states. In this case, we know that the
states are nonrelativistic states of $N_c$ quarks, each with two spins states
and $f$
flavors states (for $f$ flavors), completely symmetric in spin, flavor and
space variables.  This is conveniently described in a $(2f)^{N_c}$ dimensional
tensor product space
with independent spin indices and flavor indices for each of the quarks
(labeled by $x=$ 1 to $N_c$). In the ground states, all
the quarks will be in the same $s$-wave space wave function, and thus the
states are completely symmetric in spin and flavor. The spin and flavor states
are then described by a tower of (spin,flavor) representations of increasing
spin as shown below (for odd $N_c$):
{\renewcommand{\arraystretch}{3}
\begin{equation}\begin{array}{c}
\displaystyle\left(
\beginpicture\setcoordinatesystem units <\unitlength,\unitlength>
\put {\square} at 0 0
\put {\square} at 10 0
\put {\square} at 40 0
\put {\square} at 10 -10
\put {\square} at 40 -10
\put {$\cdots$} at 25 0
\put {$\cdots$} at 25 -10
\put {$\displaystyle\overbrace{\beginpicture\setcoordinatesystem units
<\unitlength,\unitlength>
\linethickness=0pt\putrule from 0 0 to 40 0
\endpicture}^{(N_c-1)/2}$} at 25 20
\putrule from 15 5 to 35 5
\putrule from 15 -5 to 35 -5
\putrule from 15 -15 to 35 -15
\endpicture\;,\;
\beginpicture\setcoordinatesystem units <\unitlength,\unitlength>
\put {\square} at 0 0
\put {\square} at 10 0
\put {\square} at 40 0
\put {\square} at 10 -10
\put {\square} at 40 -10
\put {$\cdots$} at 25 0
\put {$\cdots$} at 25 -10
\put {$\displaystyle\overbrace{\beginpicture\setcoordinatesystem units
<\unitlength,\unitlength>
\linethickness=0pt\putrule from 0 0 to 40 0
\endpicture}^{(N_c-1)/2}$} at 25 20
\putrule from 15 5 to 35 5
\putrule from 15 -5 to 35 -5
\putrule from 15 -15 to 35 -15
\endpicture
\right)
\quad\quad
\left(
\beginpicture\setcoordinatesystem units <\unitlength,\unitlength>
\put {\square} at 0 0
\put {\square} at 10 0
\put {\square} at -10 0
\put {\square} at -20 0
\put {\square} at 40 0
\put {\square} at 10 -10
\put {\square} at 40 -10
\put {$\cdots$} at 25 0
\put {$\cdots$} at 25 -10
\put {$\displaystyle\overbrace{\beginpicture\setcoordinatesystem units
<\unitlength,\unitlength>
\linethickness=0pt\putrule from 0 0 to 40 0
\endpicture}^{(N_c-3)/2}$} at 25 20
\putrule from 15 5 to 35 5
\putrule from 15 -5 to 35 -5
\putrule from 15 -15 to 35 -15
\endpicture\;,\;
\beginpicture\setcoordinatesystem units <\unitlength,\unitlength>
\put {\square} at 0 0
\put {\square} at 10 0
\put {\square} at -10 0
\put {\square} at -20 0
\put {\square} at 40 0
\put {\square} at 10 -10
\put {\square} at 40 -10
\put {$\cdots$} at 25 0
\put {$\cdots$} at 25 -10
\put {$\displaystyle\overbrace{\beginpicture\setcoordinatesystem units
<\unitlength,\unitlength>
\linethickness=0pt\putrule from 0 0 to 40 0
\endpicture}^{(N_c-3)/2}$} at 25 20
\putrule from 15 5 to 35 5
\putrule from 15 -5 to 35 -5
\putrule from 15 -15 to 35 -15
\endpicture
\right)
\\ \displaystyle
\left(
\beginpicture\setcoordinatesystem units <\unitlength,\unitlength>
\put {\square} at 0 0
\put {\square} at 10 0
\put {\square} at -10 0
\put {\square} at -20 0
\put {\square} at -30 0
\put {\square} at -40 0
\put {\square} at 40 0
\put {\square} at 10 -10
\put {\square} at 40 -10
\put {$\cdots$} at 25 0
\put {$\cdots$} at 25 -10
\put {$\displaystyle\overbrace{\beginpicture\setcoordinatesystem units
<\unitlength,\unitlength>
\linethickness=0pt\putrule from 0 0 to 40 0
\endpicture}^{(N_c-5)/2}$} at 25 20
\putrule from 15 5 to 35 5
\putrule from 15 -5 to 35 -5
\putrule from 15 -15 to 35 -15
\endpicture\;,\;
\beginpicture\setcoordinatesystem units <\unitlength,\unitlength>
\put {\square} at 0 0
\put {\square} at 10 0
\put {\square} at -10 0
\put {\square} at -20 0
\put {\square} at -30 0
\put {\square} at -40 0
\put {\square} at 40 0
\put {\square} at 10 -10
\put {\square} at 40 -10
\put {$\cdots$} at 25 0
\put {$\cdots$} at 25 -10
\put {$\displaystyle\overbrace{\beginpicture\setcoordinatesystem units
<\unitlength,\unitlength>
\linethickness=0pt\putrule from 0 0 to 40 0
\endpicture}^{(N_c-5)/2}$} at 25 20
\putrule from 15 5 to 35 5
\putrule from 15 -5 to 35 -5
\putrule from 15 -15 to 35 -15
\endpicture
\right)
\quad\quad \cdots
\end{array}
\label{young1}
\end{equation}}

If the quarks are not heavy, we cannot in general ignore the spin dependent
interactions.  We do not know how to write down the Hartree potential.
Nevertheless, we can plausibly argue that the low lying baryon states can be
described in the same space, and by the same representations. For this to be
the case, it is sufficient that no dramatic phase changes occur as we decrease
the quark masses from large values, much greater than the QCD scale parameter
$\Lambda$, down to small values, less than or of the order of $\Lambda$. This
certainly appears to be true for $\n3$, where the baryon octet and decuplet
bound states of the light $u$, $d$ and $s$ quarks correspond exactly to the
states we would expect if the quarks were heavy. It would be quite bizarre if
\lnc\ worked very differently.\footnote{This argument is independent of the
Hartree approximation.} In what follows, we will assume that this is
correct and see what we can say about the states for \lnc.

\subsection*{Energies}

We can write the energy of the baryon states in the Hartree language as
follows:
\begin{equation}
\sum_{n=1}^{N_c} H^n
\label{e1}
\end{equation}
where $n$ labels the number of quarks involved. It is convenient to think of
this as a matrix in the $(2f)^{N_c}\times(2f)^{N_c}$ dimensional
spin-flavor space of the baryon states. The $H^n$ are matrix functionals of
the Hartree wave function, $\Phi(\vec r)$, and it's adjoint, $\Phi(\vec
r)^\dagger$, where $\Phi$ is a
$2f\times2f$ matrix in spin and flavor space. The explicit expression involves
a tensor product of $n$ $\Phi$s and $n$ $\Phi^\dagger$s, one in each quark
space,
\begin{equation}\begin{array}{c}
\displaystyle H^n=\sum_{\{x_1,\cdots,x_n\}\atop\in\{1,\cdots,N_c\}}\int
d^3r_{x_1}\cdots d^3r_{x_n}\\
\displaystyle \Phi(\vec r_{x_1})^\dagger_{x_1}\otimes
\cdots\otimes\Phi(\vec r_{x_n})_{x_n}^\dagger
\;\tilde h^n(\vec r_{x_1},\cdots,\vec r_{x_n}) \;\Phi(\vec
r_{x_1})_{x_1}\otimes\cdots\otimes\Phi(\vec r_{x_n})_{x_n}
\end{array}
\label{e2}
\end{equation}
where the $\tilde h^n$s are $(2f)^{N_c}\times(2f)^{N_c}$ matrices acting on
the spin and flavor space, $\{x_1,\cdots,x_n\}$ refers to a set of distinct
quark lines ($x_i\neq x_j$) and $\Phi(\vec r_x)_x$ is the Hartree wave
function as a matrix acting in the space of the $x$ quark. Because each term
in the sum in (\ref{e2})
describes an interaction that involves only the $n$ quarks in the set
$\{x_1,\cdots,x_n\}$, the matrix is nontrivial only in a
$(2f)^{n}\times(2f)^{n}$ subspace (a different one for each term in the sum).
The matrix is the identity on the quark variables that are not in the set
$\{x_1,\cdots,x_n\}$. In this note, we will discuss the limit in which the $f$
light quarks are degenerate, in which case there is an $SU(f)$ flavor
symmetry, and the $\tilde h^n$s are trivial in flavor space for all of the
quarks, simply the product of identity matrices in each of the $f$ dimensional
flavor spaces of the individual quark lines. We will return to the question of
$SU(f)$ breaking in a future publication.

The trick that allows us to demonstrate the spin independence of the low-spin
states is to break up the energy according to the transformation properties of
the $\tilde h^n$ under the spin and orbital angular momentum generators on the
various quark lines in the frame of reference in which the baryon is at rest.
In particular, we will take as our ``zeroth-order'' term
in the energy, the sum of the pieces of $\tilde h^n$ that are singlets under
spin and orbital rotations, so that
\begin{equation}
\tilde h^n(\vec r_{x_1},\cdots,\vec r_{x_n})
=h^n_0(r_{x_1},\cdots,r_{x_n})+
\Delta\tilde h^n(\vec r_{x_1},\cdots,\vec r_{x_n})
\label{e3}
\end{equation}
and
\begin{equation}\begin{array}{c}
\displaystyle
H_0=\sum_{n=1}^{N_c}\sum_{\{x_1,\cdots,x_n\}\atop\in\{1,\cdots,N_c\}}\int
d^3r_{x_1}\cdots d^3r_{x_n}\\
\displaystyle \Phi(\vec r_{x_1})^\dagger_{x_1}\otimes
\cdots\otimes\Phi(\vec r_{x_n})_{x_n}^\dagger
\; h^n_0(r_{x_1},\cdots,r_{x_n}) \;\Phi(\vec
r_{x_1})_{x_1}\otimes\cdots\otimes\Phi(\vec r_{x_n})_{x_n}
\end{array}
\label{e4}
\end{equation}
where the $h_0$s are $c$-number functions (if you prefer, multiplied by the
identity matrix in the full space) of the magnitudes of the coordinates,
$r_x\equiv|\vec r_x|$. The remainder, $\Delta\tilde h^n(\vec
r_{x_1},\cdots,\vec r_{x_n})$, transforms nontrivially under the spin and/or
orbital angular momentum on at least one of the quark lines.

The plan now is to imagine using $H_0$ to determine the Hartree potential, and
then treat the $\Delta\tilde h^n(\vec r_{x_1},\cdots,\vec r_{x_n})$ as a
perturbation. Note that the $h_0^n$ for $n>1$ contribute in \lnc\ even though
they are of order $1/N_c^{n-1}$, because combinatoric factors from the sum in
(\ref{e2}) cancel the $N_c$ dependence of the individual graphs.~\cite{witten}
While we cannot determine the Hartree potential for light quarks
explicitly, even in this approximation, it is clear that resulting
ground-state
Hartree wave functions will be spin independent and functions only of $r_x$,
so that the matrix functions, $\Phi(\vec r_x)$ become $c$-numbers:
\begin{equation}
\Phi(\vec r_x)\rightarrow\phi(r_x)\,.
\label{cnumbers}
\end{equation}
In this approximation, all the spin states in the ground-state baryon
multiplet are degenerate and have the same space wave function. We can now put
the $s$-wave, spin independent Hartree wave functions back into the full
expression for the energy and ask what is the effect if the spin dependent
terms. Roughly speaking, we will find that while the terms in $H_0$ involving
two or more quarks add coherently to give a large effect, the spin dependent
terms add incoherently for the low spin states. This is why the spin dependent
terms have a small effect.

\subsection*{Splittings}

Let us now consider in detail the splittings introduced by $\Delta\tilde
h^n(\vec
r_{x_1},\cdots,\vec r_{x_n})$. We will need three simple facts:
\begin{enumerate}
\item $\Delta\tilde h^n(\vec r_{x_1},\cdots,\vec r_{x_n})$ is of order
$1/N_c^{(n-1)}$;\label{f1}
\item Terms transforming nontrivially under orbital angular momentum on any
quark line vanish when integrated over space in (\ref{e2});\label{f2}
\item The spin matrix structure on each quark line (labeled by $x$) can always
be written in the form $a+\vec b\cdot\vec\sigma_x$, where $\vec\sigma_x$ are
the Pauli matrices acting on the quark line.\label{f3}
\end{enumerate}
Using these facts, the integrations in (\ref{e2}) can be formally done, and
the results replaced by unknown constants, so that energy has the following
form (still a matrix in the $(2f)^{N_c}\times(2f)^{N_c}$ spin-flavor space):
\begin{equation}
N_cE_0+\sum_{n=1}^{N_c}\;{1\over
N_c^{n-1}}\;\sum_{\{x_1,\cdots,x_n\}\atop\in\{1,\cdots,N_c\}}
\sum_{{a_{x_1},\cdots,a_{x_n}\atop=0}}^3\;
\sigma^{a_{x_1}}_{x_1}\cdots\sigma^{a_{x_n}}_{x_n}\;
k^{a_{x_1}\cdots a_{x_n}}\,,
\label{e5}
\end{equation}
where we have defined $\sigma^0_x\equiv I$. Note that the tensor, $k$ is
completely symmetric.

The next step is note that all the terms proportional to $\sigma^0_x$ on any
quark line, $x$, are actually irrelevant. For these terms, we can do the sum
over $x$ explicitly, picking up a factor of order $N_c$. The result looks just
like the term in (\ref{e5}) with one less quark line ($n\rightarrow n-1$).
Thus if we eliminate all the $\sigma^0_x$s, it simply changes the values of
the unknown parameters in (\ref{e5}), and we can write the energy as
\begin{equation}
N_cE_0+\sum_{n=2}^{N_c}\;{1\over
N_c^{n-1}}\;\sum_{\{x_1,\cdots,x_n\}\atop\in\{1,\cdots,N_c\}}
\sum_{{a_{x_1},\cdots,a_{x_n}\atop=1}}^3\;
\sigma^{a_{x_1}}_{x_1}\cdots\sigma^{a_{x_n}}_{x_n}\;
k^{a_{x_1}\cdots a_{x_n}}\,.
\label{e6}
\end{equation}
Note that the $n=1$ term in the sum has disappeared because of rotation
invariance. There is no way to build a spin-dependent term with only one
$\vec\sigma$ in the $s$-wave ground state.

Finally, consider what happens if instead of summing over the sets
$\{x_1,\cdots,x_n\}$, we sum independently over the individual quark lines.
This introduces combinatoric factors, but for small $n$, they are of order 1
as $N_c\rightarrow\infty$ and can be absorbed into the unknown coefficients.
We also make errors by including contributions when two or more quarks lines
in the sum are the same, but these are always down by powers of $N_c$ compared
to contributions we keep because they involve fewer sums over quark lines.
Thus we can write the energy as
\begin{equation}\begin{array}{c}
\displaystyle N_cE_0+\sum_{n=2}^{N_c}\;{1\over
N_c^{n-1}}\;\sum_{x_1,\cdots,x_n\atop=1}^{N_c}
\sum_{{a_{x_1},\cdots,a_{x_n}\atop=1}}^3\;
\sigma^{a_{x_1}}_{x_1}\cdots\sigma^{a_{x_n}}_{x_n}\;
k^{a_{x_1}\cdots a_{x_n}}\\\displaystyle =N_cE_0+
N_c\sum_{n=2}^{N_c}
\sum_{{a_{1},\cdots,a_{n}\atop=1}}^3\;
{2\over N_c}S^{a_{1}}_{1}\cdots{2\over N_c}S^{a_{n}}_{n}\;
k^{a_{1}\cdots a_{n}}= N_c\,F(S^2/N_c^2)\,,
\end{array}
\label{e7}
\end{equation}
where $\vec S$ is the total spin of the baryon. Though we derived this result
thinking about first order perturbation theory in $\Delta\tilde h^n(\vec
r_{x_1},\cdots,\vec r_{x_n})$, it is actually completely general --- it is
clear that the counting of powers of $N_c$ in higher order terms goes exactly
the same way.

(\ref{e7}) has the property promised in the introduction. The spin dependent
corrections are small (order $1/N_c$ --- two factors of $N_c$ smaller than the
total energy) for fixed spin as $N_c\rightarrow\infty$. However, near the top
of the ground-state multiplet where $\vec S={\cal O}(N_c/2)$, the corrections
are as large as the zeroth order term. For the top of the multiplet, the
perturbation theory breaks down completely and the Hartree wave function for
the high spin states is not simply related to $\phi(r)$.

\subsection*{Matrix Elements}

All of the above, we believe, is well known to workers in the field, although
we have not seen it expressed in this way in the literature. The advantage of
the systematic approach described above is that we can now apply the same
ideas to discuss matrix elements of operators between baryon states. The
results will be similar to those for the baryon energies. We will find \lnc\
predictions for the matrix elements that are corrected by terms that are
smaller by a factor of order $S^2/N_c^2$. Thus the predictions should be
reliable at the bottom of the ground state baryon multiplet.

Consider first the matrix elements of two-quark operators of the form
\begin{equation}
\overline{\psi}\;\Gamma\;\psi
\label{m1}
\end{equation}
where $\Gamma$ is a product of a flavor matrix $\Lambda$ times some $\gamma$
matrix. In a \lnc\ baryon state, the matrix element will involve the
same flavor matrix, $\Lambda$, and a matrix, $\kappa$, in spin space that is
either 1 or $\vec\sigma$ depending on whether the matrix element is a scalar
or a vector under rotations. There may also be orbital contributions to the
spin structure of the operator, but these are irrelevant to matrix elements in
the low-spin ground states because the expectation values are computed in the
$s$-wave Hartree wave functions. Thus the form of the operator on any single
quark line, $x$, is proportional to $\Lambda_x\kappa_x$, where $\Lambda_x$ and
$\kappa_x$ are just the matrices $\Lambda$ and $\kappa$ acting on the flavor
and spin spaces of the $x$ quark.

Arguments precisely analogous to those that we used to discuss the energies
then imply that the leading order contribution to the matrix element of the
operator (which like the energy, we will express as a
$(2f)^{N_c}\times(2f)^{N_c}$ matrix in the spin-flavor space) has the form:
\begin{equation}
\sum_{x=1}^{N_c}\;\left(
a\Lambda_x\kappa_x+b\sigma^a_x\Lambda_x\kappa_x\sigma^a_x\right)
\,,
\label{m2}
\end{equation}
where $a$ and $b$ are constants. The second term in (\ref{m2}) arises because
there are effects that are leading order in $N_c$ from diagrams of the form
shown in Fig.~\ref{annulus}, where the shaded region represents some planar
collection of gluon lines. This term does not have a dramatic effect on the
form of the matrix element, but it allows for the possibility that vector and
scalar operators are renormalized differently. Except for this unknown
difference in normalization of vector and scalar operators, (\ref{m2}) is the
standard result of $SU(2f)$ symmetry arguments. Note also that the constants
$a$ and $b$ will depend on the details of the $\gamma$ matrix structure in
(\ref{m1}). For example, the space component of the axial vector current, with
$\Gamma=\Lambda\vec\gamma$ will yield different values of $a$ and $b$ than the
operator with $\Gamma=\Lambda\sigma^{ab}$, even though both transform like
vectors. As noted by Dashen and Manohar~\cite{dandm}, the matrix elements
(\ref{m2}) can be of order $N_c$ because of the sum over quarks.

\begin{figure}[htb]
$$\beginpicture
\setcoordinatesystem units <\tdim,\tdim>
\putrule from -75 0 to 75 0
\setdots <\tdim>
\circulararc 180 degrees from 25 0 center at 0 0
\circulararc 180 degrees from 10 0 center at 0 0
\setshadegrid span <\tdim>
\vshade -25 0 0 <,z,,> -24 0 7 <,z,,> -20 0 15 <,z,,> -15 0 20
<,z,,> -10 0 23 <,z,,> -8 6 23 <,z,,> -7 7 24
<,z,,> -6 8 24
<,z,,> 0 10 25 <,z,,> 6 8 24
<,z,,> 7 7 24 <,z,,> 8 6 23 <,z,,> 10 0 23 <,z,,> 15 0 20 <,z,,> 20 0 15
<,z,,> 24 0 7
<,z,,> 25 0 0 /
\put {$\times$} at 0 0
\linethickness=0pt
\putrule from 0 -10 to 0 35
\endpicture$$
\caption{\label{annulus} General diagram contributing to renormalization
of two-quark operators.}
\end{figure}

Our result, (\ref{m2}), contains the results of Dashen and
Manohar~\cite{dandm} and Jenkins~\cite{jenkins} for the isovector axial
vector current, but it is more general. For example, it predicts similar
$SU(6)$ relations for the isoscalar axial vector current as well.

\subsection*{Four-quark operators}

Next consider matrix elements of four-quark operators, such as
\begin{equation}
\overline{\psi}\;\Gamma^1\;\psi\;\;
\overline{\psi}\;\Gamma^2\;\psi
\label{fm1}
\end{equation}
As above, the matrix element in \lnc\ baryons will involve the substitution
\begin{equation}
\Gamma^1\rightarrow\Lambda^1\kappa^1\,,\quad\quad
\Gamma^2\rightarrow\Lambda^2\kappa^2\,,
\label{fm1s}
\end{equation}
where the $\kappa^j$s are $\sigma^j$s or identity matrices in spin space
depending on whether $\Gamma^j$ is a vector or scalar under rotations.

Now by the usual argument, the leading \lnc\ prediction for the matrix
elements in the ground-state baryon states is
\begin{equation}\begin{array}{c}\displaystyle
\sum_{x_1,x_2\atop=1}^{N_c}
\biggl(a_1\Lambda^1_{x_1}\kappa^1_{x_1}\Lambda^2_{x_2}\kappa^2_{x_2}
+a_2\sigma^a_{x_1}\Lambda^1_{x_1}\kappa^1_{x_1}\sigma^a_{x_1}
\Lambda^2_{x_2}\kappa^2_{x_2}\\ \displaystyle
+a_3\Lambda^1_{x_1}\kappa^1_{x_1}
\sigma^a_{x_2}\Lambda^2_{x_2}\kappa^2_{x_2}\sigma^a_{x_2}
+a_4\sigma^a_{x_1}\Lambda^1_{x_1}\kappa^1_{x_1}\sigma^a_{x_1}
\sigma^a_{x_2}\Lambda^2_{x_2}\kappa^2_{x_2}\sigma^a_{x_2}
\biggr)\,,
\end{array}
\label{fm2}
\end{equation}
which, because of the double sum, can grow like $N_c^2$ as
$N_c\rightarrow\infty$.

It should be clear to the reader how similar predictions can be obtained for
matrix elements of operators with more quark fields.

\subsection*{Comments}

It is worth restating the warning in the introduction about the application of
these predictions to $N_c=3$. In our view, the only sensible way to proceed is
to take the result (\ref{m2}) or (\ref{fm2}) and apply it for $\n3$, because
real \lnc\
baryons simply do not look anything like $\n3$ baryons. Unfortunately, this is
not always done. For example, calculations in Skyrmion models~\cite{skyrmion}
are not consistent with this view. A particularly obvious
problem with Skyrmion calculation is that they yield non-zero proton
matrix-elements of $\overline{s}s$ operators. It is clear that these matrix
elements are non-zero for \lnc\ for precisely the same reason that the \lnc\
baryons and the $\n3$ baryons have very different quantum numbers. The strange
quarks in the \lnc\ baryon are not part of the sea. They are valence quarks!
The $\overline{s}s$ operators have nonzero matrix elements in \lnc\ baryons
simply because the \lnc\ baryons have the wrong valence structure. Our
prediction for matrix elements of $\overline{s}s$ operators, based on
(\ref{m2}) or (\ref{fm2}) or their generalizations, is zero to leading order
in $N_c$, as you should expect from the
absence of quark loops in leading order \lnc\ calculations.

In this paper, we have set up a formalism that is useful for exploiting the
approximate spin independence of low spin baryon states in a systematic way.
In a future publication, we will give some examples of applications of
(\ref{m2}) and (\ref{fm2}) and discuss the effects of $SU(f)$ flavor symmetry
breaking. We will also give explicit examples of $1/N_c$ corrections and show
how to apply these arguments to excited baryon states.

\section*{Acknowledgements} HG is grateful for useful conversations with Roger
Dashen, Bob Jaffe, David Kaplan, Aneesh Manohar and Ann Nelson.

\end{document}